**Finite-sample bias-variance tradeoff with variables related to trial participation inserted into causal forest models for ensuring generalizability**

Rikuta Hamaya, MD, PhD, MS[1], Etsuji Suzuki, MD, PhD[2], Konan Hara, MD, PhD[3]

[1] Division of Preventive Medicine, Department of Medicine, Brigham and Women's Hospital and Harvard Medical School, Boston, MA

[2] Department of Epidemiology, Graduate School of Medicine, Dentistry and Pharmaceutical Sciences, Okayama University, Okayama, Japan

[3] Harris School of Public Policy, University of Chicago, Chicago, IL

**Address for correspondence:**

Rikuta Hamaya, MD, PhD, MS

Division of Preventive Medicine, Brigham and Women's Hospital

900 Commonwealth Avenue East, Boston, MA 02215

Tel.: +1 617 278 0835

Fax: +1 617 731 3843

E-mail: rhamaya@bwh.harvard.edu

**Funding:** This work is funded by the Evergreen Innovation Fund.

**Conflict of interests:** RH holds stock and is a scientific advisor of Everyone Cohort Inc.

**Abstract**

Estimating conditional average treatment effects (CATE) from randomized controlled trials (RCTs) and generalizing them to broader populations is essential for personalizing treatment rules but is complicated by selection bias due to trial participation and potentially high-dimensional covariates. We evaluated finite-sample bias-variance tradeoff for Causal Forest–based CATE estimation strategies to address the selection bias. Identification theory suggests unbiased CATE estimation is possible when covariates related to trial participation are included in CATE estimating models. However, simulation studies demonstrated that, under realistic RCT sample sizes, variance inflation from high-dimensional covariates often outweighed modest bias reduction. In our data-generating process that define individual treatment effect (ITE) in source population and selected trial samples, including >3 covariates related to participation in causal forest substantially degraded precision unless sample sizes were large. In contrast, inverse probability weighting (IPW)-based methods consistently improved performance across scenarios. Application to a RCT of omega-3 fatty acids and coronary heart disease illustrated how IPW shifts CATE estimates toward source population effects and refines heterogeneity assessments. Our findings highlight that including trial-selection variables for CATE estimating models may inflate estimator variance and reduce ITE prediction performance in applications using medical RCTs. Addressing selection bias separately (e.g. through IPW) would be a reasonable strategy.

**Key messages**

• Estimating conditional average treatment effects (CATE) with machine learning methods such as Causal Forest allows incorporation of multiple covariates, but generalizing results from RCTs to the source population raises concerns about selection bias and variance inflation.

• Our finite-sample simulations showed that including trial participation covariates yields unbiased CATE estimates in the source population in theory, but the increase in variance often outweighs the reduction in bias, worsening overall performance.

• Integrating inverse probability weighting to handle selection bias improved the generalizability of CATE estimates in our data generating process.

## Introduction

Moving beyond the average treatment effect (ATE), estimating the conditional average treatment effect (CATE) has been gaining attention in medical and epidemiological research. Conventionally, CATE is estimated by conditioning on single or a few sets of covariates, often using non-parametric models. Recently, machine-learning (ML)-based approaches to estimate CATE with many covariates have been proposed. These include meta-learners[1–3] and Causal Forest[4,5], and their application may contribute to personalized medicine through considering more covariates, or more finely predicting individual treatment effect (ITE)[6,7] than conventional approaches do. Compared with one-size-fits-all approaches based on ATE, CATE-based treatment rules, or formally individualized treatment rules (ITRs)[8,9], could potentially maximize the benefits of treatments while minimizing the harms by selectively identifying the individuals who are indicated for the intervention. Several application studies are emerging[10–12], especially based on well-conducted randomized controlled trials (RCTs) data.

Although there have been lots of literatures about the approaches to estimate CATE for multiple covariates or in high-dimension settings, the generalizability is not always assured. For example, we previously showed that some meta-learner algorithms properly work with high-dimensional covariates including demographics, lifestyle, and genetic information in a dietary trial[13], while the CATE estimates may not be necessary generalizable to general populations[14]. Assuring generalizability is particularly relevant when CATE is derived from RCTs, because the potential selection bias through trial participation may influence the performance of CATEs in a practical setting beyond trial samples[15–17]. There are several approaches to account for generalizability which may be applied to either ATE[16,18] or CATE[19].

This paper systematically evaluates the finite-sample bias-variance tradeoffs of these strategies, comparing the direct inclusion of trial-participation covariates within Causal Forests and inverse probability weighting (IPW)-based adjustment. We hypothesized that direct inclusion would suffers from severe variance inflation under realistic sample sizes of clinical trials.

## Identification

### *Settings*

We use an RCT setting for outcome $Y$ (dichotomous or continuous) in which a dichotomous treatment $A$ (1 = treated, 0 = untreated) is marginally randomized within the trial. We let $Y(a)$ denote a counterfactual outcome of $Y$ had the treatment $A$, possibly contrary to the fact, been set to $a$. We also let $S$ denote a dichotomous selection variable and consider that the participants in the RCT ($S = 1$), or trial sample, are selected from the source population (i.e. nested trial design[17]). We define source population as a sufficiently large pool of individuals who are potentially indicated for the intervention, including trial non-participants ($S = 0$). **Figure 1** illustrates the present setting in a directed acyclic graph (DAG). We consider $\boldsymbol{X_1}$ being covariates of interest for conditioning on CATE and $\boldsymbol{X_2}$ as other determinants of $S$ than (parts of) $\boldsymbol{X_1}$ ($\boldsymbol{X_1} \cap \boldsymbol{X_2} = \emptyset$). Note that $\boldsymbol{X_2}$ may or may modify the effect of $A$ on $Y$. We assume that $\boldsymbol{X_1}$ and $\boldsymbol{X_2}$ are observed in the source population (i.e. $S = 0$ or 1), and $Y$ and $A$ are observed only in trial samples ($S = 1$). There can be unmeasured covariates $U$ that modify the effects of $A$ on $Y$, so unbiasedly estimating ITE is not possible. For simplicity, we assume no loss to follow-up, no measurement error, and perfect adherence of the assigned treatment. We do not consider variables observed after treatment is initiated, including mediators, since the present aim is to identify who would benefit from the treatment at baseline. Although **Figure 1** indicates that $\boldsymbol{X_1}$ and $\boldsymbol{X_2}$ are direct causes of $Y$, the following arguments hold when these are not direct causes but associated with $Y$.[20]

We consider an absolute difference scale and set the following two aims:

Aim A) to estimate $CATE(\boldsymbol{x_1}) = \mathrm{E}[Y(1) - Y(0) | \boldsymbol{X_1} = \boldsymbol{x_1}]$

Aim B) to estimate $CATE(\boldsymbol{x_1}, \boldsymbol{x_2}) = \mathrm{E}[Y(1) - Y(0) | \boldsymbol{X_1} = \boldsymbol{x_1}, \boldsymbol{X_2} = \boldsymbol{x_2}]$

In Aim A), we are interested in CATE given a specific covariate or combinations of covariates; $\boldsymbol{X_1}$ may thus either single or a few covariates, several covariates that are often prespecified in an RCT protocol for subgroup analysis, or many covariates such as genetic information that make the setting high-dimensional. In Aim B), we are interested in predicting ITE as closely as possible. As we cannot consistently estimate ITE in the presence of *U,* we may want to use (selected or comprehensive) observed effect modifiers $\boldsymbol{X_1}$ as well as other variables determining trial participation $\boldsymbol{X_2}$, aiming to model more individualized effects in the source population. Thus, caution is needed for $\boldsymbol{X_1}$ being potentially differently defined according to the Aims.

***Assumptions***

To identify these CATEs, we further assume the following conditions[17,19]:

*1) Conditional exchangeability over treatment in the trial:*

$Y(a) \perp\!\!\!\perp A|\boldsymbol{X_1}=\boldsymbol{x_1}, \boldsymbol{X_2}=\boldsymbol{x_2}, S=1$ for each $a$ and every $\boldsymbol{x_1}$ and $\boldsymbol{x_2}$ with positive density in the trial $f(\boldsymbol{x_1}, \boldsymbol{x_2}, S=1)>0$.

*2) Positivity of treatment in the trial:*

$\Pr[A=a|\boldsymbol{X_1}=\boldsymbol{x_1}, \boldsymbol{X_2}=\boldsymbol{x_2}, S=1]>0$ for each $a$ and every $\boldsymbol{x_1}$ and $\boldsymbol{x_2}$ with positive density in the trial $f(\boldsymbol{x_1}, \boldsymbol{x_2}, S=1)>0$.

*3) Conditional exchangeability over trial participation:*

$Y(a) \perp\!\!\!\perp S|\boldsymbol{X_1}=\boldsymbol{x_1}, \boldsymbol{X_2}=\boldsymbol{x_2}$ for each $a$ and every $\boldsymbol{x_1}$ and $\boldsymbol{x_2}$ with positive density in the trial $f(\boldsymbol{x_1}, \boldsymbol{x_2}, S=1)>0$.

*4) Positivity of trial participation:*

$\Pr[S=1|\ \boldsymbol{X_1}=\boldsymbol{x_1}, \boldsymbol{X_2}=\boldsymbol{x_2}]>0$ for every $\boldsymbol{x_1}$ and $\boldsymbol{x_2}$ with positive density in the trial $f(\boldsymbol{x_1}, \boldsymbol{x_2}, S=1)>0$.

*5) Consistency of potential outcomes in the trial:*

For all trial participants ($S=1$), $Y=Y(a)$ if $A=a$.

***Estimation approaches***

Researchers generally hope to estimate $CATE(\boldsymbol{x_1})$ or $CATE(\boldsymbol{x_1}, \boldsymbol{x_2})$ in the source population (i.e. $S=0$), and ensuring the generalizability is especially important in developing an ITR for clinical practice, i.e. beyond RCT settings. In theory, with RCT data only, these CATE estimands in the source population can be consistently estimated if we condition on the covariates including those determining trial participation (i.e. $\boldsymbol{X_1}$ and $\boldsymbol{X_2}$).[15,17] Formal identifications are available in **Appendix 1**.

We explain three distinct approaches to estimate $CATE(\boldsymbol{x_1})$ and $CATE(\boldsymbol{x_1}, \boldsymbol{x_2})$ (to predict ITE), aiming to achieve a greater prediction performance that takes both bias and variance into account, such as mean squared error (MSE). We use Causal Forest as an example of CATE-estimating algorithms that can accommodate many covariates.

*Naïve approach (Model 1): fit Causal Forest using effect modifiers of interests in trial samples*

When we ignore the selection bias due to trial participation, we may be only interested in $\boldsymbol{X_1}$. This obviously leads to a biased estimate for either $CATE(\boldsymbol{x_1})$ or $CATE(\boldsymbol{x_1}, \boldsymbol{x_2})$, while the estimation process is simplest and the variance can be smaller. This is particularly true when the dimension of $\boldsymbol{X_1}$ is low or contributions of $\boldsymbol{X_2}$ to $CATE(\boldsymbol{x_1}, \boldsymbol{x_2})$ are smaller than those of $\boldsymbol{X_1}$; in these cases, CATE estimation through Causal Forest can lead to pronouncedly smaller variances when using $\boldsymbol{X_1}$ only compared with using both $\boldsymbol{X_1}$ and $\boldsymbol{X_2}$.

*CATE algorithms to account for selection bias (Model 2): fit Causal Forest using effect modifiers of interests and determinants of trial participation in trial samples*

When we are concerned about selection bias, we may easily account for it by just including the relevant covariates (i.e. $\boldsymbol{X_1}$ and $\boldsymbol{X_2}$) in the Causal Forest algorithm. This simple consideration gives us a consistent estimator for Aim B. For Aim A, we obtain a consistent estimator by averaging the contributions of $\boldsymbol{X_2}$ to estimate $CATE(\boldsymbol{x_1})$, which can be achieved using Monte Carlo integration for example; after fitting Causal Forest in trial samples, we draw many samples of $\boldsymbol{x_2}$ from its conditional probability distribution $p(\boldsymbol{x_2}|\boldsymbol{x_1})$ from source population data, and then we average estimated $\mathrm{E}[Y(1) - Y(0)|\boldsymbol{X_1} = \boldsymbol{x_1}, \boldsymbol{X_2} = \boldsymbol{x_2}, S = 1]$. Or, this can be parametrically estimated using a sample analog like g-formula[16]. A drawback of approach is high dimensionality in the CATE estimation, which can inflate variances.

*IPW to account for selection bias (Models $1_{IPW}$ and $2_{IPW}$): fit Causal Forest using effect modifiers of interests in IPW-weighted trial samples*

Another way to handle selection bias is to use IPW through estimating probability of trial participation using $\boldsymbol{X_2}$ in the source population data. In the weighted samples, we need to decide which covariates to include in Causal Forest, i.e. $\boldsymbol{X_1}$ only (Model $1_{IPW}$) or $\boldsymbol{X_1}$ and $\boldsymbol{X_2}$ (Model $2_{IPW}$). For Aim A, the benefit of Model $1_{IPW}$ compared with Model 1 is clear and it should lead to smaller bias as long as $Pr(S = 1|\boldsymbol{X_1} = \boldsymbol{x_1}, \boldsymbol{X_2} = \boldsymbol{x_2})$ is correctly specified. Model $2_{IPW}$ needs Monte Carlo integration as in Model 2. For Aim B, Model $1_{IPW}$ is biased but may perform better than Model $2_{IPW}$ depending on the extent of variance inflation due to high dimensionality in Causal Forest. Meanwhile, Model $2_{IPW}$ can be considered 'doubly robust' for selection bias in a sense that it is consistently estimated if either IPW or Causal Forest model is correct.

## Simulation

We aimed to evaluate how including different sets of covariates in a Causal Forest model influences the prediction performance of $CATE(\boldsymbol{X_1})$ (Aim A) and ITE through $CATE(\boldsymbol{X_1}, \boldsymbol{X_2})$ estimation (Aim B) in the source population. For Aim A, we hypothesized that, compared with Causal Forest including $\boldsymbol{X_1}$ only (i.e. using a subset of target effect modifiers only; Model 1), including both $\boldsymbol{X_1}$and $\boldsymbol{X_2}$ (i.e. using both a subset of target effect modifiers and variables determining trial participation; Model 2) would reduce bias for $CATE(\boldsymbol{X_1})$ by addressing selection bias, while increase variance due to higher dimension, leading to the prediction performance (as measured by MSE) not necessarily better. In addition, we hypothesized that Causal Forest including $\boldsymbol{X_1}$ only (Model $1_{IPW}$) or $\boldsymbol{X_1}$ and $\boldsymbol{X_2}$ (Model $2_{IPW}$) in the IPW-weighted trial samples could perform better than other approaches in some scenarios. For Aim B, including $\boldsymbol{X_1}$ (i.e. all target effect modifiers) and $\boldsymbol{X_2}$ would be expected to perform better than including $\boldsymbol{X_1}$ only for ITE prediction, since it covers effect modifiers more comprehensively, whereas a similar bias-variance tradeoff would also be expected depending on the strength of effect modification by $\boldsymbol{X_2}$ and/or dimension of $\boldsymbol{X_1}$. We acknowledge that there is another approach using pseudo-outcomes[19], but this approach cannot be extended to high-dimensional CATE in its present from, and thus we did not compare this approach to Causal Forest in the present simulation.

### *Data generation*

Lists of covariates $\boldsymbol{X_{1ALL}}$ (all effect modifiers of interest), $\boldsymbol{X_2}$ (covariates determining trial participation), and $\boldsymbol{O}$ (other unobserved effect modifiers) are independently generated from standard normal distributions, with the dimensions of 20, 10, and 20 total covariates, respectively. The simulation assumes that only $\boldsymbol{X_2}$ determine trial participation for simplicity; this assumption is trivial because we always adjust $\boldsymbol{X_1}$ regardless of their contributions to $S$. The true ITE is defined as: $ITE = 1 * \boldsymbol{X_{1ALL}} + \boldsymbol{0.5} * \boldsymbol{X_2} + 0.3 * \boldsymbol{O}$. We also examined the case where the coefficients on $\boldsymbol{X_2}$ is 0 (i.e. no contribution of $\boldsymbol{X_2}$ in treatment effect), as the generalizability does not require selection correction for the trial participation in this case.

$\boldsymbol{X_1}$ (effect modifiers of interest for a particular study) is a subset of $\boldsymbol{X_{1ALL}}$ with the number of covariates being either 2, 3, 5, or 10. The above ITE definition follows $CATE(\boldsymbol{X_1}) = 1 * \boldsymbol{X_1}$, because the CATE integrates out the variables other than $\boldsymbol{X_1}$ in the ITE, which are independent of $\boldsymbol{X_1}$. We consider dichotomous treatment $A$ and continuous outcome $Y$. Treatment $A$ is assigned completely at random with a probability 0.5 in the trial. $Y$ is defined with $\epsilon$ (random noise from a normal distribution with mean 0 and variance 1) such that $Y = \boldsymbol{X_{1ALL}} + \boldsymbol{X_2} + \boldsymbol{O} + A * ITE + \epsilon$. Of note, treatment status $A$ in the source population is not generated, as only ITE is needed for validating estimated CATE in the source population.

Without replacement, we selected trial samples from the source population (N = 100,000), with the probability of being included in the trial depends only on $\boldsymbol{X_2}$: $P[S = 1|\boldsymbol{X_2}] = \mu * expit(1 * \boldsymbol{X_2} + 0.2 * \boldsymbol{X_2}^2)$, where $\mu$ is a scaling factor such that the target trial sample size is achieved. Target trial sample was set as either 200, 500, 2,000, or 5,000.

### *Simulation Approaches*

We aim to estimate $CATE(\boldsymbol{x_1})$ (i.e. Aim A) and predict $ITE$ through estimating $CATE(\boldsymbol{x_1}, \boldsymbol{x_2})$ (i.e. Aim B) in the source population using trial data and covariate information on source population. We train separate Causal Forest models on the trial data using $\boldsymbol{X_1}$ (potential effect modifiers) only as Model 1, and $\boldsymbol{X_1}$ and $\boldsymbol{X_2}$ (variables determining trial participation) as Model 2. Additionally, we calculate probabilities to be selected into the trial using a logistic regression model based on $\boldsymbol{X_2}$ linearly, and we train Causal Forest models in the resulting IPW-weighted trial samples using $\boldsymbol{X_1}$ only (Model $1_{\text{IPW}}$) and $\boldsymbol{X_1}$ and $\boldsymbol{X_2}$ (Model $2_{\text{IPW}}$). Note that the logistic regression model for trial participation is misspecified because we do not include quadratic terms for $\boldsymbol{X_2}$.

For Aim a, in Models 2 and $2_{\text{IPW}}$, we estimate $CATE(\boldsymbol{X_1})$ from estimated $CATE(\boldsymbol{X_1}, \boldsymbol{X_2})$ through the Monte Carlo integration approximation with respect to $p(\boldsymbol{x_2}|\boldsymbol{x_1})$; given the independence between $\boldsymbol{X_1}$ and $\boldsymbol{X_2}$, we sampled directly from the empirical marginal distribution of $\boldsymbol{X_2}$ for deriving $p(\boldsymbol{x_2}|\boldsymbol{x_1})$. We repeat the whole analysis for 50 times to evaluate the performance of CATE estimation. The primary evaluation metrics are mean squared error (MSE) for actual $CATE(\boldsymbol{X_1})$ and ITE in the source population in Aims A and B, respectively. We additionally calculate bias and variances (as MSE minus bias-squared) for each simulation. All analyses are conducted using R version 4.2.0 (The R Foundation).

***Results***

***Aim A: Estimation of*** $\boldsymbol{CATE(x_1)}$

**Figure 2** summarizes the evaluation metrics of $CATE(\boldsymbol{x_1})$ in the source population (y-axis), separately based on four different models, according to the sample size of the trial (x-axis) and numbers of continuous covariates $\boldsymbol{X_1}$ for CATE (different panels). Note that, comparing evaluation metrics across different $\boldsymbol{X_1}$ is not possible, because the estimation targets are different.

Model 1 had highest bias with trial sample size ≥500, arising from the selection bias due to trial participation. Model 2 had slightly smaller bias compared with Model 1 through CATE estimation including $\boldsymbol{X_2}$, while IPW-based selection bias adjustment in Models $1_{\text{IPW}}$ and $2_{\text{IPW}}$ led to much quicker bias convergence to 0. Influence due to misspecification in the logistic regression models was smaller (i.e. through comparing Model $1_{\text{IPW}}$ [misspecified model] versus Model $2_{\text{IPW}}$ [doubly robust]). Most of the variance was determined by the number of covariates in Causal Forest model, leading to smaller variance in Models 1 and $1_{\text{IPW}}$ compared with Models 2 and $2_{\text{IPW}}$ for $\boldsymbol{X_1}$ with ≥3 covariates. For $\boldsymbol{X_1}$ with 2 covariates, variance of Models 2 and $2_{\text{IPW}}$ got smaller than that of Models 1 and $1_{\text{IPW}}$ with trial sample size larger than ~1,500. Variance is much larger than bias when $\boldsymbol{X_1}$ with ≥3 covariates is considered. Consequently, MSE of Model 2 is smaller than that of Model 1 when $\boldsymbol{X_1}$ with 2 covariates is considered and trial sample size larger than ~500. For $\boldsymbol{X_1}$ with ≥3 covariates, much larger sample size is needed for better performance in Model 2. Performance gain through IPW-adjustment is observed in almost every setting. When there is no contribution of $\boldsymbol{X_2}$ in treatment effects (**eFigure 1**), there was no bias in any models with sample size >~2,000. Variances of each model were same as the main simulation.

The observation highlights that, when considering multiple covariates in Causal Forest to address selection bias, much larger sample size would be necessary than practical medical RCTs. In other words, even when we consider only several covariates to estimate CATE, which we may typically consider ‘low-dimensional’, imprecise prediction due to inflated variance may cancel out the benefit in reduced bias when we concurrently input $\boldsymbol{X_2}$ in Causal Forest. It is also worth noting that Model $1_{IPW}$ had similar-to-better performance than Model 1 in most settings, even in the cases where there is no selection bias or Model 1 is correct. In scenarios resembling our data generating process, where selection mechanisms are linearly approximated and dimensionality is moderately high, employing IPW adjustment offered a favorable bias-variance compromise and may be preferable to direct inclusion of covariates in the forest.

***Aim B: Estimation of*** $\boldsymbol{CATE(x_1, x_2)}$

Results for ITE evaluation through $CATE(\boldsymbol{x_1}, \boldsymbol{x_2})$ estimation are summarized in **Figure 3.** Findings for bias were similar to those in Aim A. Variances were generally much greater than in those in Aim a, reflecting more covariates contributing to ITE than $CATE(\boldsymbol{x_1})$. At the same time, more covariates in Causal Forest (i.e. greater number of $\boldsymbol{X_1}$) led to smaller variance due to less model specifications. Variance in Model 2 was smaller than in Model 1 when $\boldsymbol{X_1}$ with 2 covariates is considered and trial sample size larger than ~500, and when $\boldsymbol{X_1}$ with 3 covariates is considered and trial sample size larger than ~2,500. Consequently, MSE was smaller in Model 2 than Model 1 for when $\boldsymbol{X_1}$ with 2 covariates is considered and trial sample size larger than ~300, and when $\boldsymbol{X_1}$ with 3 covariates is considered and trial sample size larger than ~1,500. However, with more $\boldsymbol{X_1}$, performance was better in Model 1. IPW adjustment led to smaller MSE in most settings.

When there is no contribution of $\boldsymbol{X_2}$ in treatment effects (**eFigure 1**), there was no bias in any models with sample size >~2,000, similarly in Aim a. Variances were smaller in Models 1 and 3, as including $\boldsymbol{X_2}$ do not improve the $CATE(\boldsymbol{X_1})$ estimation but increase noises due to the trial participation probability estimation. As a result, Model 2 only performed better than Model 1 when $\boldsymbol{X_1}$ with 2 covariates is considered and trial sample size larger than ~2500.

In our fine-sample data generating process, even when aiming to estimate ITE through estimating CATE for many variables, reduction in bias by including more covariates may not benefit as much as increase in variance due to higher dimensionality in Causal Forest. Reliably predicting ITE (i.e. CATE for many covariates) may not be thus achieved by naively using a Causal Forest model.

## Application example

We provide an application using the VITAL study (ClinicalTrials.gov: NCT01169259) to investigate CATE of omega-3 FA supplementation on coronary heart disease (CHD) incidence in 25,871 US older adults (men aged ≥50 years and women aged ≥55 years)[21,22]. Briefly, VITAL is a two-by-two factorial, double-blind, placebo-controlled testing omega-3 FA of 840 mg/day (1.2:1 ratio of eicosapentaenoic acid [EPA] to docosahexaenoic acid [DHA]) and vitamin D3 (at a dose of 2000 IU per day) in the primary prevention of cardiovascular disease and cancer. The VITAL is so far the only completed primary prevention RCT of omega-3 FA supplementation, which showed non-significant reduction of the primary endpoint of major cardiovascular disease, a composite of myocardial infarction (MI), stroke, or death from cardiovascular causes, while it observed a significant 17% reduction in total CHD[21]. In addition, there are much heterogeneity in the treatment effect according to baseline covariates[21]. In the present application example, we compared the performance of CATE for different sets of covariates regarding effects of omega-3 FA supplementation on incident CHD, given that omega-3 is expected to prevent CHD[23]. National Health and Nutrition Examination Survey (NHANES)[24] 2013-2014 data, applied to VITAL inclusion criteria, was considered as the source population. Details in the approaches are available in **Appendix 2**.

***Results***

**eTable 1** shows the distribution of covariates in NHANES and VITAL data at baseline. **Figure 4** illustrates the distributions of estimated $CATE(\boldsymbol{x_1})$ and $CATE(\boldsymbol{x_1}, \boldsymbol{x_2})$ in the source population, according to each Model. The distributions centered at the ATE, which was higher in Models 1 and 2 compared with IPW-adjusted models. For Aim A, Monte Carlo integration changed the distributions of estimated CATE. For Aim B, including $\boldsymbol{X_1}$ and $\boldsymbol{X_2}$ in Causal Forest increased the ranges of estimated CATE compared with that in $\boldsymbol{X_1}$ only.

Experimental evaluation results of CATE in trial samples are summarized in **eFigure 3**. Compared with ATE of -0.61% (95% CI: -1.27, 0.05) absolute risk change of CHD by omega-3 FA supplementation, effects in lowest tertiles of estimated CATE in each model (those with high expected benefit) had slightly greater CHD risk reduction. However, the confidence intervals are wider and not much heterogeneity in the treatment effect could be captured. In addition, unlike simulation, we do not have either outcome data in the source population or counterfactual outcomes in trial samples, and thus caution is needed for the evaluation.

## Discussion

We investigated different approaches aiming to estimate CATE for multiple covariates that is generalizable to the source population. Identifiable condition suggests that including covariates determining trial selection in CATE-estimating models like Causal Forest can yield unbiased CATE estimate in the source population. Our finite-sample simulation showed that even when we consider only a few covariates to estimate CATE through this approach, the harm due to inflated variance may cancel out the benefit in reduced bias, either intending to estimate CATE for specific covariates or to estimate ITE. CATE estimation had greater performance in most settings when selection bias was addressed through IPW. Our findings highlight that including trial-selection variables for CATE estimating models may worsen performance in typical medical trials (e.g., N < 5000) when the selection variables weakly modify the treatment effect or when the added dimensionality heavily inflates estimator variance. Addressing selection bias separately (e.g. through IPW) would be a reasonable strategy.

To theoretically ground these empirical observations, we can frame them within the bias-variance tradeoff inherent to non-parametric, tree-based estimators[25]. The MSE of a causal forest asymptotically scales at a rate proportional to $\mathcal{O}\left(n^{-c/(c+p)}\right)$, where n is the sample size, p is the dimension of the feature space, and c is a parameter reflecting the smoothness of the true underlying CATE function[4,26]. Finite-sample variance inflation by including trial participation variables ($\boldsymbol{X_2}$) into the forest increases is most likely to dominate the asymptotic reduction in bias under the following conditions typical of our simulation and medical trials: (1) when $\boldsymbol{X_2}$ are weak modifiers of the treatment effect, leading to the absolute reduction in bias negligible; (2) when the expanded dimensionality is large relative to sample size, which degrades the convergence rate; and (3) when the inclusion of $\boldsymbol{X_2}$ dilutes the random subspace feature-sampling process at each split, decreasing the probability that strong effect modifiers ($\boldsymbol{X_1}$) are evaluated as candidates, thereby forcing suboptimal tree construction.

When considering which covariates to use for CATE-estimating models, differentiating $\boldsymbol{X_1}$ and $\boldsymbol{X_2}$ may provide a guidance; Using only $\boldsymbol{X_1}$ for CATE estimating models may minimize the variance of estimated CATEs, and selection bias can be addressed by using $\boldsymbol{X_1}$ and $\boldsymbol{X_2}$ for a model to derive IPW. Indeed, $\boldsymbol{X_1}$ and $\boldsymbol{X_2}$ can substantially differ; covariates expected to modify the main effect are often chosen based on underlying mechanisms and are usually prespecified in RCTs for subgroup analyses. Selecting covariates from such variables would be useful to estimate CATE allowing a larger inter-individual variability. In contrast, variables determining trial participation are not always very apparent in typical medical RCT datasets. Previous studies have addressed such variables by comparing the trial participants with external data that is representative of the source population[27]. Studies also showed that reasons of trial participation are more related to altruism[28,29], socioeconomic status[30], and structural barriers due to race[31]. These variables may need to be accounted for to achieve exchangeability over trial selection, and it is important to design an RCT to collect such data.

There are several limitations of this study. First, our argument is based on simulations in a limited context. Although we designed them such that they resemble in an actual application, and an extensive set of sensitivity analyses provided consistent findings, the findings cannot be generalized in practical settings. Still, realizing the bias-variance tradeoff is important when intending to estimate source-population CATEs through causal forest. Second, we could not compare the performance with a previously proposed method to transport CATE estimate[19], as it is not readily extendable to high-dimensional CATE. Such extensions are highly warranted in future studies. Third, the empirical application could not allow evaluation of CATE in the source population. Such evaluations may be infeasible in almost all real-world settings, and thus we believe that demonstrating how it is applied would be still informative.

## **Conclusions**

Including covariates that determine trial participation in CATE-estimating algorithms theoretically yields unbiased CATE in the source population. However, when using Causal Forest for finite samples, the prediction performance was not necessarily greater than CATE with limited covariates due to variance inflation. Separately accounting for selection bias, e.g. with IPW, would be a reasonable strategy.

**Acknowledgements**

We appreciate Sarah E Robertson (CAUSALab, Harvard T.H. Chan School of Public Health, Boston, MA) for her feedback on an earlier version of this article.

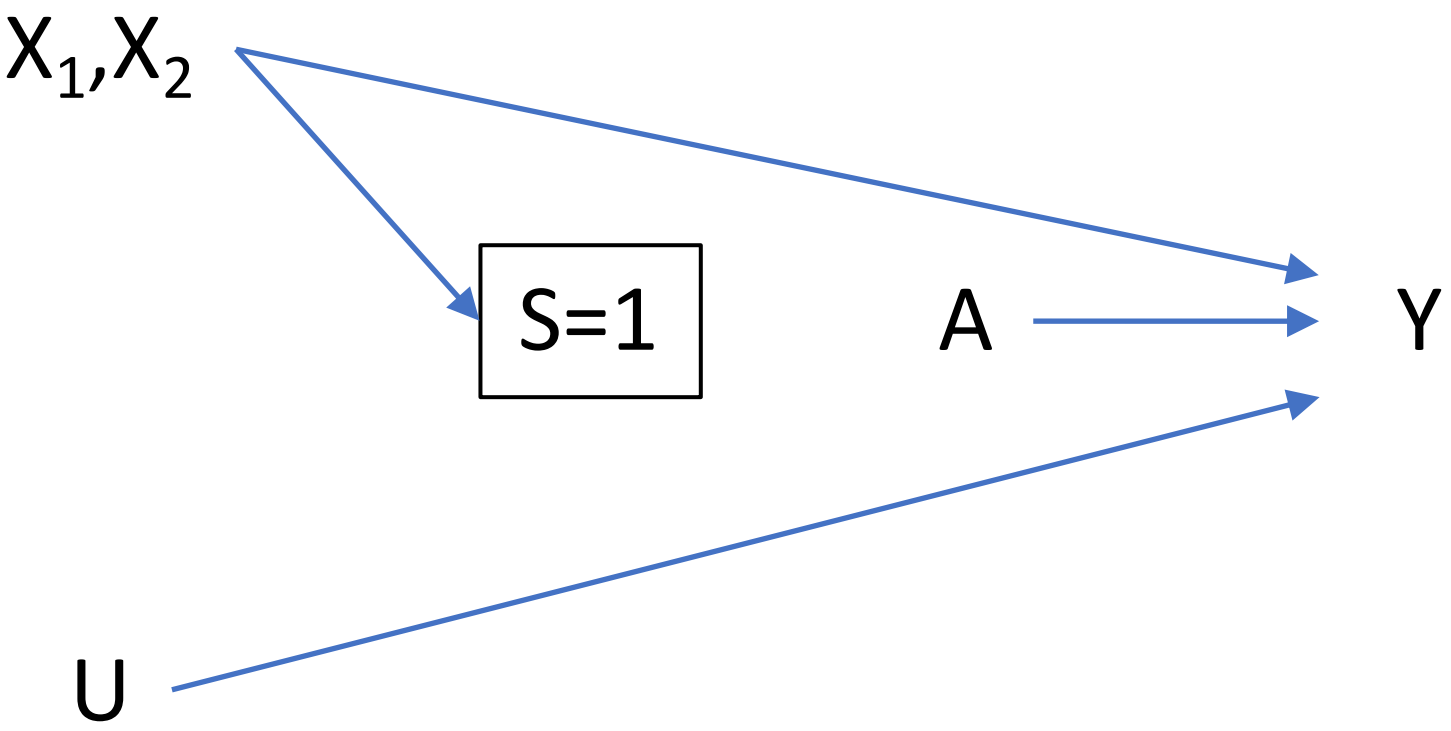

**Figure 1: A directed acyclic graph (DAG) for randomized participants**

In a randomized controlled trial ($S$ = 1), we aim to estimate the effects of treatment $A$ on outcome $Y$ conditional on measured effect modifiers $X_1$ and $X_2$. $X_1$ is a set of effect modifiers of interest, while $X_1$ and $X_2$ determine trial participation $S$. We let $O$ denote unmeasured other effect modifiers. There can be unmeasured confounders $U$ between the effect modifiers and $Y$. The DAG does not apply to the source population as the intervention is not randomized.

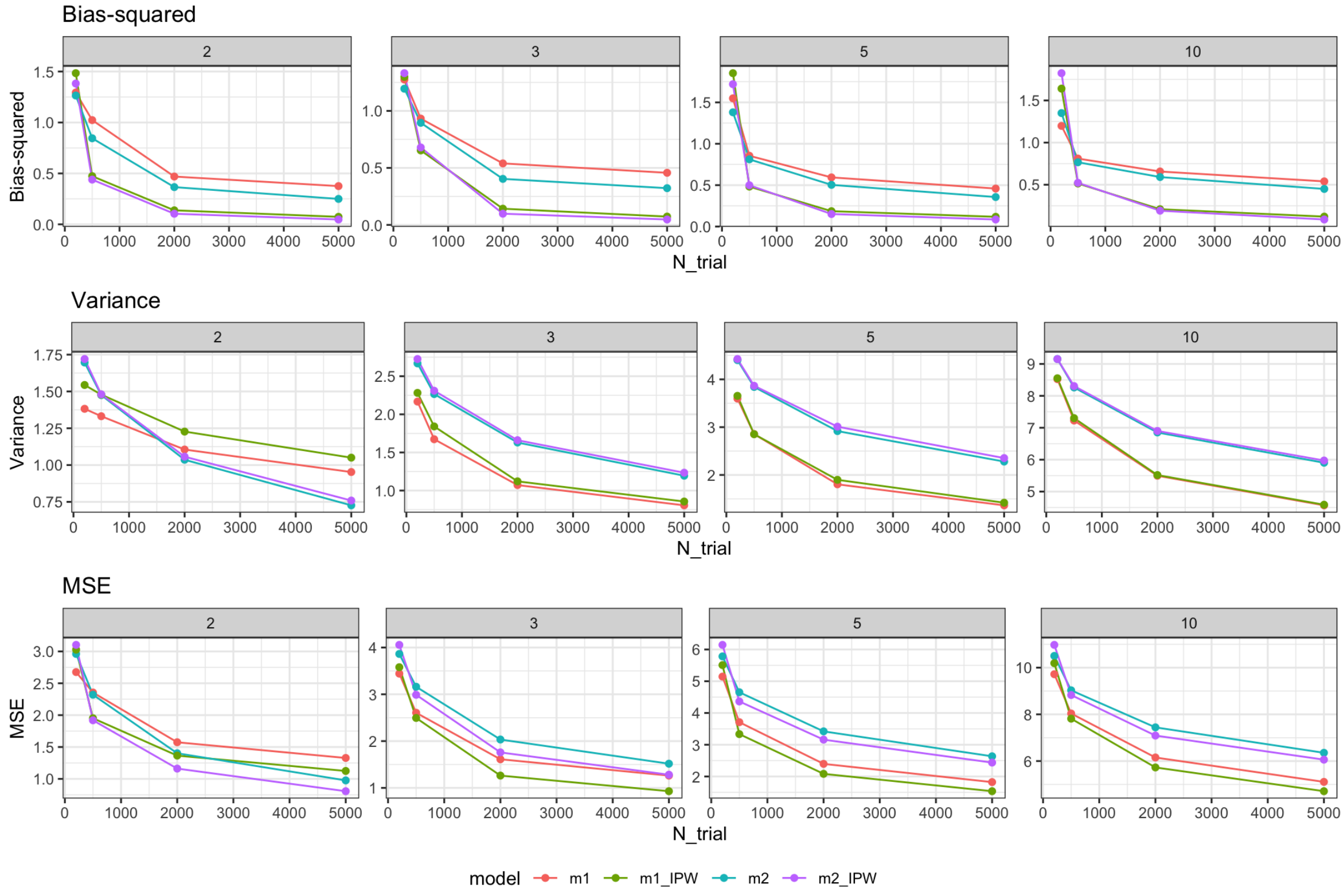

**Figure 2: Evaluation of $CATE(x_1)$**

Evaluation metrics of $CATE(x_1)$ in the source population (y-axis) when it is estimated in a randomized controlled trial with varying sample size (x-axis) and number of $\boldsymbol{X_1}$ (numbers shown above the figures). Colors describe different approaches: trained Causal Forest using $\boldsymbol{X_1}$ only (model 1), $\boldsymbol{X_1}$ and $\boldsymbol{X_2}$ (model 2), $\boldsymbol{X_1}$ only in weighted trial samples to adjust for selection bias (model $1_{IPW}$), and $\boldsymbol{X_1}$ and $\boldsymbol{X_2}$ in weighted trial samples to adjust for selection bias (model $2_{IPW}$). Monte Carlo integration was applied in models 2 and $2_{IPW}$ to estimate $CATE(x_1)$. Contribution of $\boldsymbol{X_2}$ in modifying effect was assumed to be half of that of $\boldsymbol{X_1}$.

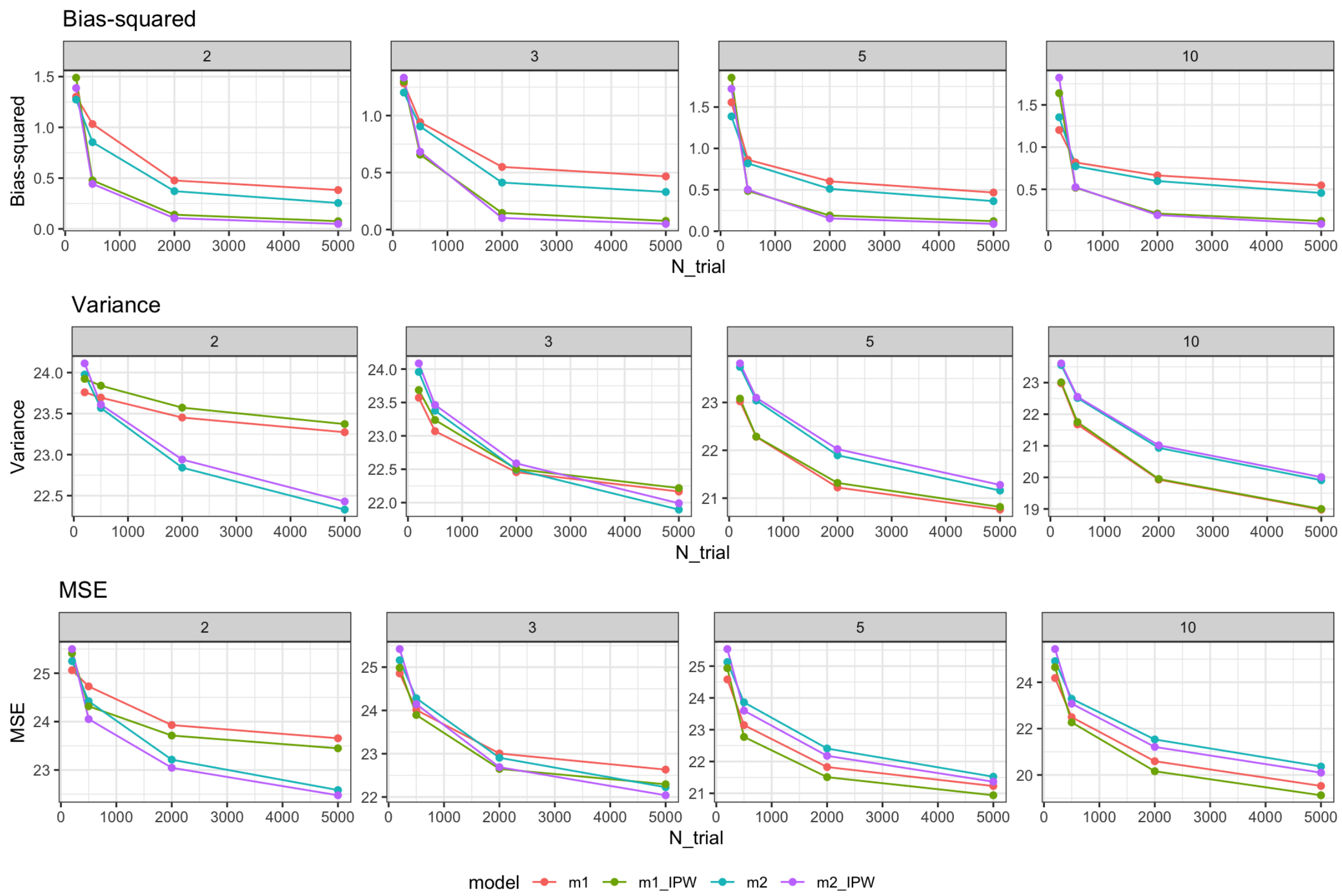

**Figure 3: Evaluation of $ITE$**

Evaluation metrics of $ITE$ in the source population (y-axis) when it is estimated in a randomized controlled trial with varying sample size (x-axis) and number of $\boldsymbol{X_1}$ (numbers shown above the figures). Colors describe different approaches: trained Causal Forest using $\boldsymbol{X_1}$ only (model 1), $\boldsymbol{X_1}$ and $\boldsymbol{X_2}$ (model 2), $\boldsymbol{X_1}$ only in weighted trial samples to adjust for selection bias (model $1_{\mathrm{IPW}}$), and $\boldsymbol{X_1}$ and $\boldsymbol{X_2}$ in weighted trial samples to adjust for selection bias (model $2_{\mathrm{IPW}}$). Contribution of $\boldsymbol{X_2}$ in modifying effect was assumed to be half of that of $\boldsymbol{X_1}$.

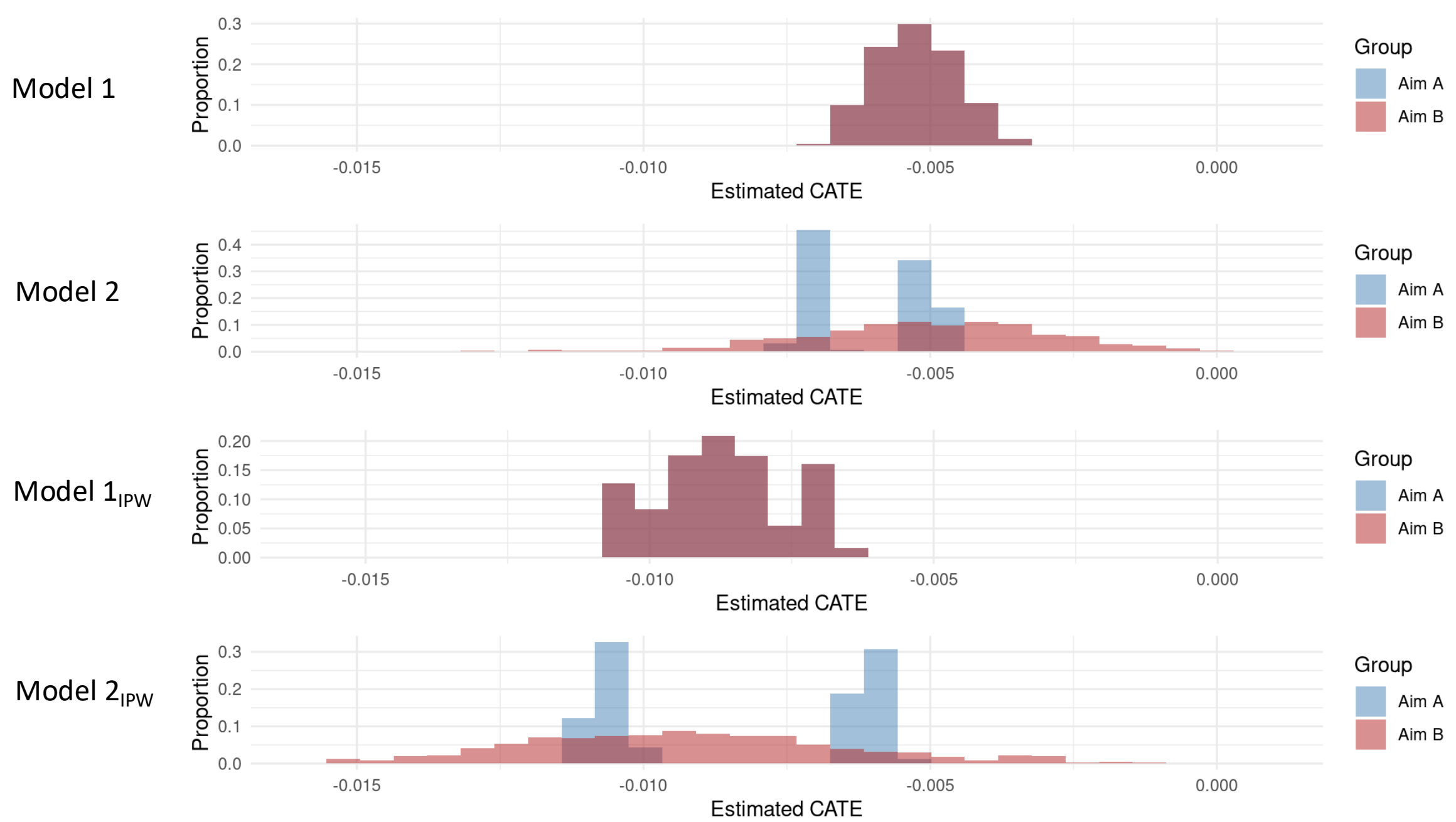

**Figure 4: Distribution of estimated CATE through Causal Forest model in VITAL trial in NHANES**

Causal Forest to estimate $CATE(x_1)$ (Aim A) and $CATE(x_1, x_2)$ (Aim B) were trained in VITAL trial via different models. Histograms show the distributions of estimated CATEs in the source population (NHANES data). For Aim A, Monte Carlo integration was applied to estimate $CATE(x_1)$ in Models 2 and $2_{\mathrm{IPW}}$.

**Supplemental Data**

Hamaya et al. Finite-sample bias-variance tradeoff with variables related to trial participation inserted into causal forest models for ensuring generalizability

**Appendix 1: Identification formulas for $\mathbf{CATE(x_1)}$ or $\mathbf{CATE(x_1, x_2)}$**

Aim A) $CATE(\boldsymbol{x_1})$

$= \mathrm{E}[Y(1) - Y(0) | \boldsymbol{X_1} = \boldsymbol{x_1}]$

$= \int \mathrm{E}[Y(1) - Y(0) | \boldsymbol{X_1} = \boldsymbol{x_1}, \boldsymbol{X_2} = \boldsymbol{x_2}]\, p(\boldsymbol{x_2} | \boldsymbol{x_1})\, d\boldsymbol{x_2}$

$= \int (\mathrm{E}[Y(1) - Y(0) | \boldsymbol{X_1} = \boldsymbol{x_1} \boldsymbol{X_2} = \boldsymbol{x_2}, S = 1]) p(\boldsymbol{x_2} | \boldsymbol{x_1})\, d\boldsymbol{x_2}$ ($\because Y(a) \perp\!\!\!\perp S | \boldsymbol{X_1}, \boldsymbol{X_2}$)

Note that $\mathrm{E}[Y(1) - Y(0) | \boldsymbol{X_1} = \boldsymbol{x_1}, \boldsymbol{X_2} = \boldsymbol{x_2}, S = 1]$ is identified under the assumptions[1,2], and $p(\boldsymbol{x_2} | \boldsymbol{x_1})$ is identified from the source population, which concludes the identification of $CATE(\boldsymbol{x_1})$. This formula, known as g-formula[3,4], suggests that we can consistently estimate $CATE(\boldsymbol{x_1})$ by integrating out $\boldsymbol{x_2}$ from estimates of $\mathrm{E}[Y(1) - Y(0) | \boldsymbol{X_1} = \boldsymbol{x_1}, \boldsymbol{X_2} = \boldsymbol{x_2}, S = 1]$, which can also be consistently estimated using ML-based algorithms that allow CATE estimation in high-dimensional settings such as Causal Forest.

It can be also estimated using Horvitz-Thompson IPW by the following equivalence:

$CATE(\boldsymbol{x_1})$

$= \int (\mathrm{E}[Y(1) - Y(0) | \boldsymbol{X_1} = \boldsymbol{x_1}, \boldsymbol{X_2} = \boldsymbol{x_2}, S = 1]) p(\boldsymbol{x_2} | \boldsymbol{x_1})\, d\boldsymbol{x_2}$

$= E[E[Y(1) - Y(0) | \boldsymbol{X_1} = \boldsymbol{x_1}, \boldsymbol{X_2} = \boldsymbol{x_2}, S = 1] | \boldsymbol{X_1} = \boldsymbol{x_1}]$

$= E[\frac{\{Y(1) - Y(0)\} * I(S=1)}{Pr(S=1 | \boldsymbol{X_1} = \boldsymbol{x_1}, \boldsymbol{X_2} = \boldsymbol{x_2})} | \boldsymbol{X_1} = \boldsymbol{x_1}]$

Similarly, we can consistently estimate this by fitting CATE-estimating algorithms using $\boldsymbol{X_1}$ as the conditioning variables in IPW-weighted trial samples, in which selection bias due to trial participation is accounted for. This approach can be advantageous in the stability of CATE estimation compared with the g-formula that needs a joint distribution of $\boldsymbol{X_1}$ and $\boldsymbol{X_2}$ in the source population**.** In turn, the estimation of the probability of trial participation can be unstable with a limited trial sample size relative to source population.

Aim B) $CATE(\boldsymbol{x_1}, \boldsymbol{x_2})$

$= \mathrm{E}[Y(1) - Y(0) | \boldsymbol{X_1} = \boldsymbol{x_1}, \boldsymbol{X_2} = \boldsymbol{x_2}]$

$= \mathrm{E}[Y(1) - Y(0) | \boldsymbol{X_1} = \boldsymbol{x_1}, \boldsymbol{X_2} = \boldsymbol{x_2}, \boldsymbol{S} = \boldsymbol{1}]$ $(\because Y(a) \perp\!\!\!\perp S | \boldsymbol{X_1}, \boldsymbol{X_2})$

This is identified under the assumptions, and its estimation simpler than Aim A – we can get consistent estimates of $CATE(\boldsymbol{x_1}, \boldsymbol{x_2})$ just by applying Causal Forest or meta-learners using both $\boldsymbol{X_1}$ and $\boldsymbol{X_2}$ in trial samples. Typically, by increasing the dimension of $\boldsymbol{X_1}$, we hope to get CATE estimates closer to ITE.

## Appendix 2: Approaches for the application example

We estimated CATE for effects of omega-3 FA supplementation on incident CHD in absolute risk scale. ATE was estimated by a linear regression model. Like the simulation, we considered the following six dichotomous covariates as $\boldsymbol{X_1}$, effect modifiers of interests: fish consumption (physiologically modify the effect of omega-3 supplementation) and age, sex, BMI, hypertension, and diabetes. We considered the following variables as $\boldsymbol{X_2}$ from race (White/ Black/ Asian/ Hispanic/Others), education level (no high school/ high school/ college/post-college), income (<$15K / $15K to <$70-75K/ ≥$70-75K; slightly different cutoffs between VITAL and NHANES), current smoking, alcohol intake (daily/ some/ none), use of aspirin and statin. Causal Forest was used to estimate CATE for $\boldsymbol{X_1}$ only (Model 1) and $\boldsymbol{X_1}$ plus $\boldsymbol{X_2}$ (Model 2), separately. Additionally, we estimated the probability of trial participation based on $\boldsymbol{X_2}$ using NHANES data, and estimated CATE using $\boldsymbol{X_1}$ only (Model $1_{IPW}$) and $\boldsymbol{X_1}$ and $\boldsymbol{X_2}$ (Model $2_{IPW}$) in the IPW-weighted trial data. For Aim A, we estimate $CATE(\boldsymbol{x_1})$ from estimated $CATE(\boldsymbol{x_1}, \boldsymbol{x_2})$ through the Monte Carlo integration approximation with respect to $p(\boldsymbol{x_2}|\boldsymbol{x_1})$ in Models 2 and $2_{IPW}$. $p(\boldsymbol{x_2}|\boldsymbol{x_1})$ was approximated by first finding similar observations in $\boldsymbol{X_1}$ space and using their corresponding $\boldsymbol{X_2}$ values through K-nearest neighbor algorithm. Distributions of the estimated $CATE(\boldsymbol{x_1})$ and $CATE(\boldsymbol{x_1}, \boldsymbol{x_2})$ in the source population were compared across each Aim and approach. Additionally, for Aim B, $CATE(\boldsymbol{x_1}, \boldsymbol{x_2})$ was estimated via cross-validation and evaluated within trial samples by estimating mean effects according to tertile of estimated CATE[1].

**References for Appendix**

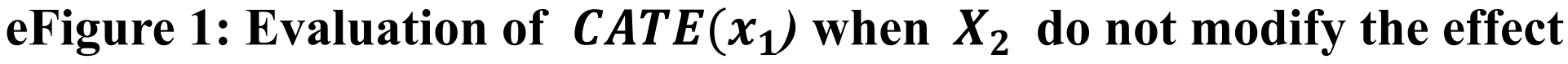

**eFigure 1: Evaluation of $CATE(x_1)$ when $X_2$ do not modify the effect**

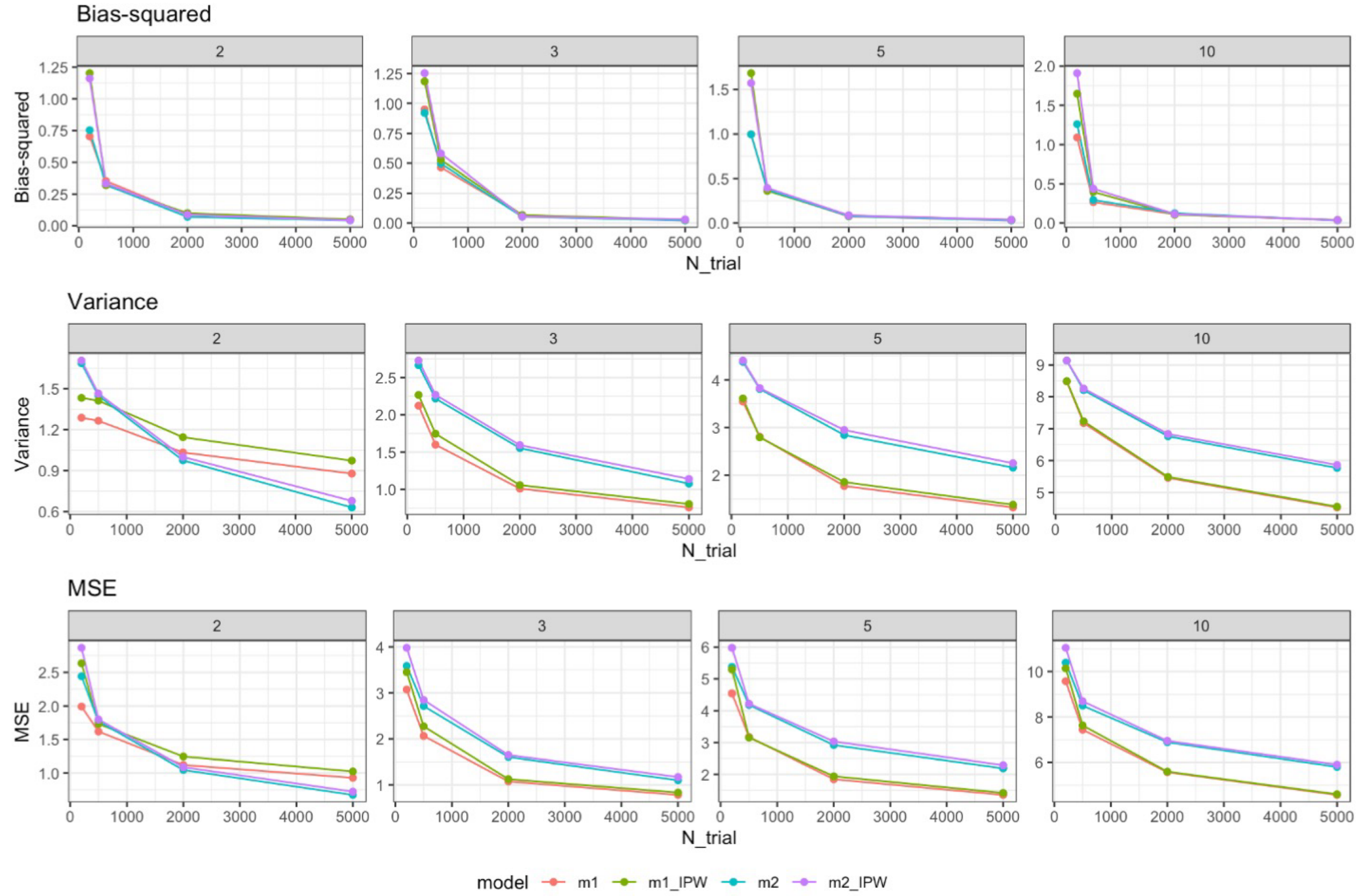

Evaluation metrics of $CATE(x_1)$ in the source population (y-axis) when it is estimated in a randomized controlled trial with varying sample size (x-axis) and number of $\boldsymbol{X_1}$ (numbers shown above the figures). Colors describe different approaches: trained Causal Forest using $\boldsymbol{X_1}$ only (model 1), $\boldsymbol{X_1}$ and $\boldsymbol{X_2}$ (model 2), $\boldsymbol{X_1}$ only in weighted trial samples to adjust for selection bias (model $1_{IPW}$), and $\boldsymbol{X_1}$ and $\boldsymbol{X_2}$ in weighted trial samples to adjust for selection bias (model $2_{IPW}$). Monte Carlo integration was applied in models 2 and $2_{IPW}$ to estimate $CATE(x_1)$. No effect modification by $\boldsymbol{X_2}$ was assumed.

**eFigure 2: Evaluation of $ITE$**

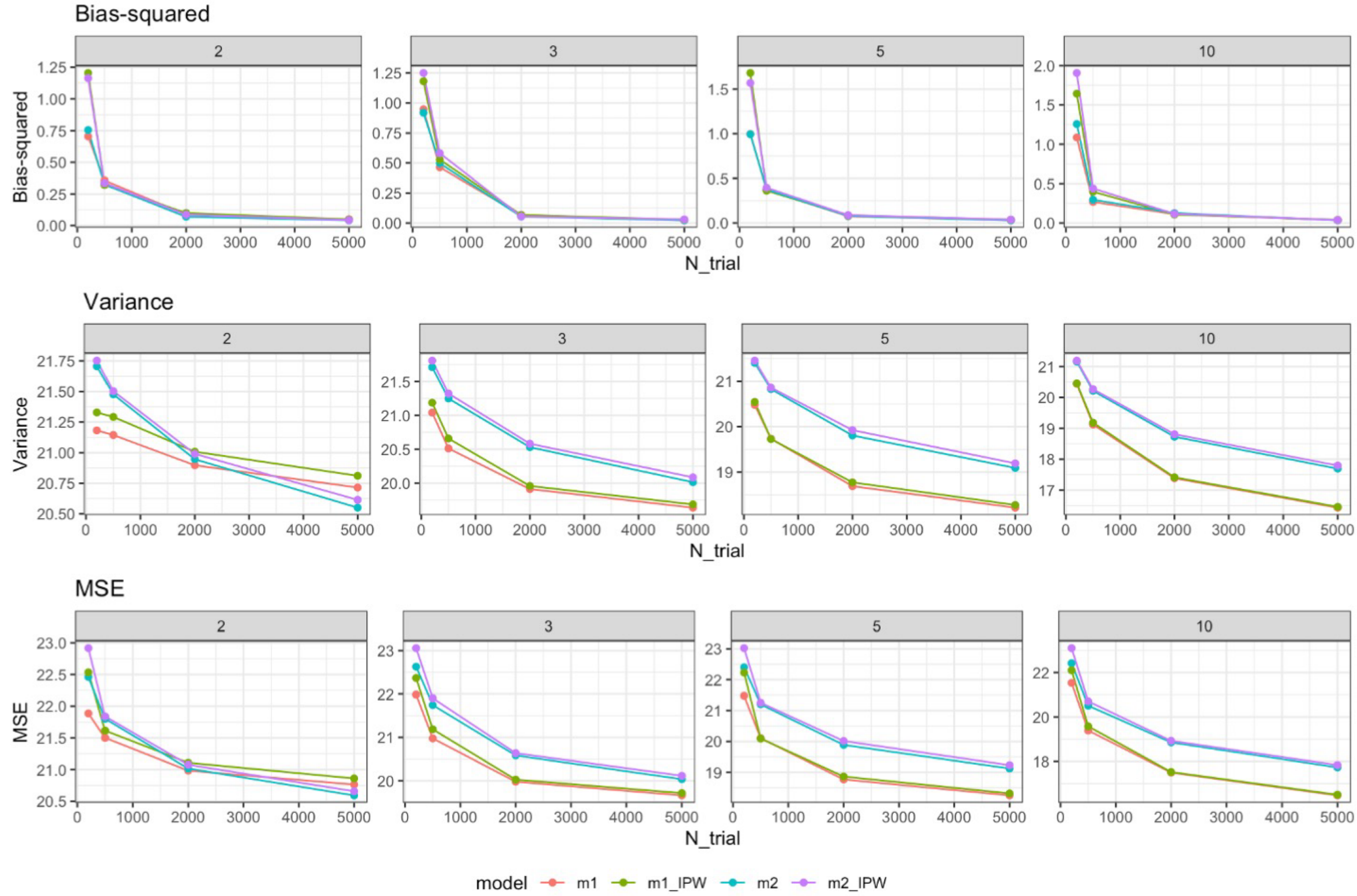

Evaluation metrics of $ITE$ in the source population (y-axis) when it is estimated in a randomized controlled trial with varying sample size (x-axis) and number of $\boldsymbol{X_1}$ (numbers shown above the figures). Colors describe different approaches: trained Causal Forest using $\boldsymbol{X_1}$ only (model 1), $\boldsymbol{X_1}$ and $\boldsymbol{X_2}$ (model 2), $\boldsymbol{X_1}$ only in weighted trial samples to adjust for selection bias (model $1_{\text{IPW}}$), and $\boldsymbol{X_1}$ and $\boldsymbol{X_2}$ in weighted trial samples to adjust for selection bias (model $2_{\text{IPW}}$). No effect modification by $\boldsymbol{X_2}$ was assumed.

**eFigure 3: Experimental evaluations of $CATE(x_1, x_2)$ in VITAL**

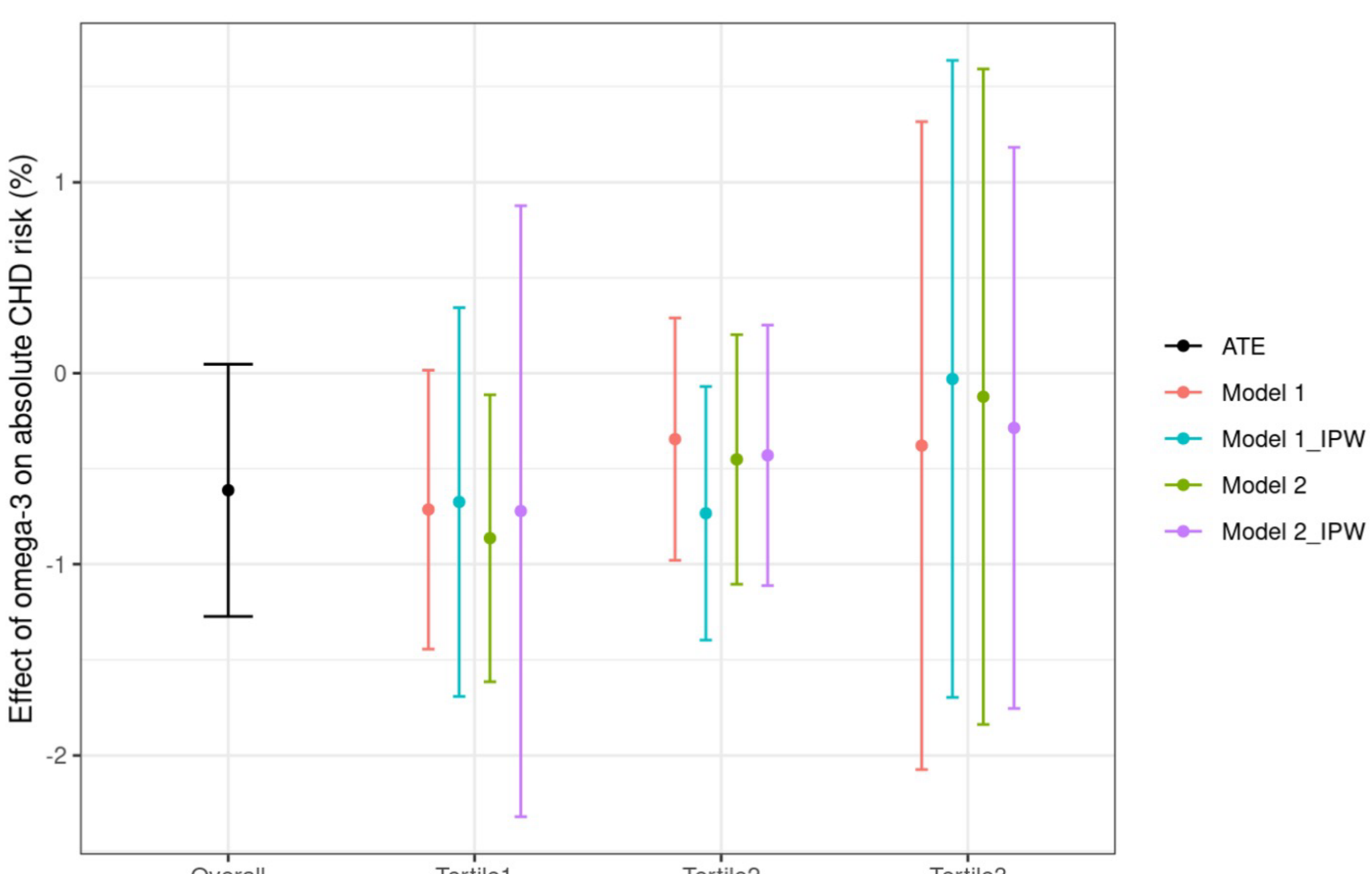

$CATE(x_1, x_2)$ was estimated via cross-validation and evaluated within VITAL trial samples by estimating mean effects according to tertile of estimated CATE. Black point shows the estimate of average treatment effect and colored points indicate estimates of the sorted, grouped average treatment effect according to estimated CATE though different models. Error bars show the 95% confidence intervals of each estimate.

eTable 1: Baseline covariate distributions in NHANES and VITAL

| | NHANES | VITAL |
|---|---|---|
| **$X_1$** | | |
| Mean age, years | 63.1 | 67.1 |
| Men | 51.7 | 49.4 |
| Mean BMI, $kg/m^2$ | 29.4 | 28.1 |
| Hypertension | 50.8 | 51.6 |
| Diabetes | 15.4 | 13.7 |
| Fish intake > 1.5 serving/day | 26.5 | 51.6 |
| **$X_2$** | | |
| White race | 71.5 | 71.9 |
| Black race | 11.2 | 19.7 |
| Hispanic race | 10.2 | 3.9 |
| Asian race | 5.6 | 1.5 |
| Other race | 1.5 | 2.9 |
| Education: Post-college | 30.5 | 45.1 |
| Education: College | 30.7 | 42 |
| Education: High school | 21.5 | 11.4 |
| Education: No high school | 17.3 | 1.4 |
| Income: <$15K | 11.9 | 6.3 |
| Income: $15K to <$70-75K | 50.9 | 46.6 |
| Income: ≥$70-75K | 37.2 | 47.1 |
| Current smoking | 14.2 | 7.1 |
| Alcohol: daily | 7.9 | 25.7 |
| Alcohol: some | 54.5 | 41.8 |
| Alcohol: none | 37.7 | 32.6 |
| Use of antihypertensives | 40.9 | 19.5 |
| Use of cholesterol-lowering drugs | 33.5 | 36.9 |
| Use of diabetes drugs | 12.1 | 10.6 |
| Use of aspirin | 33.9 | 44.7 |
| Use of statins | 33.5 | 34.4 |

Numbers are mean for age and BMI and % for other variables.
VITAL inclusion criteria have been applied to NHANES dataset.